\documentclass[fleqn,10pt]{wlscirep}
\usepackage[utf8]{inputenc}
\usepackage[T1]{fontenc}

\usepackage{amsmath}
\usepackage{amssymb}
\usepackage{amsthm}
\usepackage{graphicx}
\usepackage{lipsum}
\usepackage{authblk}
\usepackage{geometry}
\usepackage{xcolor}
\usepackage{adjustbox}
\usepackage{mathtools}
\usepackage{amsfonts}
\usepackage{relsize}
\usepackage{multirow}
\usepackage{booktabs}
\usepackage{tabularx}
\usepackage{listings}

\usepackage{xspace}
\usepackage[ruled,vlined]{algorithm2e}

\usepackage{multirow}

\newcommand{\mname}{\text{SkipGNN}\xspace}
\newcommand{\eg}{\emph{e.g.}\xspace}
\newcommand{\ie}{\emph{i.e.}\xspace}

\newtheorem*{problem*}{Problem}

\newcommand{\xhdr}[1]{\vspace{2mm}\noindent{{\bf #1.}}}

\title{SkipGNN: Predicting Molecular Interactions with Skip-Graph Networks}

\author[1]{Kexin Huang}
\author[2]{Cao Xiao}
\author[2]{Lucas M. Glass}
\author[3]{Marinka Zitnik}
\author[4,*]{Jimeng Sun}

\affil[1]{Health Data Science, Harvard T.H. Chan School of Public Health, Boston, MA}
\affil[2]{Analytic Center of Excellence, IQVIA, Cambridge, MA}
\affil[3]{Department of Biomedical Informatics, Harvard University, Boston, MA}
\affil[4]{Department of Computer Science, University of Illinois at Urbana-Champaign, Urbana, IL}

\affil[*]{jimeng@illinois.edu}

\begin{abstract}
Molecular interaction networks are powerful resources for molecular discovery. They are increasingly used with machine learning methods to predict biologically meaningful interactions. While deep learning on graphs has dramatically advanced the prediction prowess, current graph neural network (GNN) methods are {\color{black} mainly} optimized for prediction on the basis of direct similarity between interacting nodes. In biological networks, however, similarity between nodes that do not directly interact has proved incredibly useful in the last decade across a variety of interaction networks. Here, we present \mname, a graph neural network approach for the prediction of molecular interactions. \mname predicts molecular interactions by not only aggregating information from direct interactions but also from second-order interactions, which we call skip similarity. In contrast to existing GNNs, \mname receives neural messages from two-hop neighbors as well as immediate neighbors in the interaction network and non-linearly transforms the messages to obtain useful information for prediction. To inject skip similarity into a GNN, we construct a modified version of the original network, called the skip graph. We then develop an iterative fusion scheme that optimizes a GNN using both the skip graph and the original graph. Experiments on four interaction networks, including drug-drug, drug-target, protein-protein, and gene-disease interactions, show that \mname achieves superior and robust performance. Furthermore, we show that unlike popular GNNs, \mname learns biologically meaningful embeddings and performs especially well on noisy, incomplete interaction networks.
\end{abstract}
\begin{document}

\flushbottom
\maketitle

\thispagestyle{empty}

\section{Introduction}
\label{sec:intro}

Molecular interaction networks are ubiquitous in biological systems. Over the last decade, interaction networks have advanced our systems-level understanding of biology~\cite{cowen2017network}. Further, they have enabled discovery of biologically significant, yet previously unmapped relationships~\cite{zitnik2019machine}, including drug-target interactions (DTIs)~\cite{luo2017nature}, drug-drug interactions (DDIs)~\cite{zitnik2018modeling}, protein-protein interactions (PPIs)~\cite{luck2020reference}, and gene-disease interactions (GDIs)~\cite{agrawal2018large}. To assist in these discoveries, a plethora of computational methods, primarily optimized for link prediction from networks (\eg, \cite{lei2013novel}), were developed to predict new interactions in molecular networks.
Recently, deep learning on graphs has emerged as a dominant class of methods that have revolutionized state-of-the-art in learning and reasoning over network datasets. These methods, often referred to as graph neural networks (GNNs)~\cite{wu2019comprehensive} and graph convolutional networks (GCNs)~\cite{kipf2017semi,velivckovic2017graph}, operate by performing a series of non-linear transformations on the input molecular network, where each transformation aggregates information only from immediate neighbors, \ie, direct interactors in the network. While these methods yield powerful predictors, they explicitly take into account only direct similarity between nodes in the network. Therefore, GNNs are limited at fully capturing important information for prediction that resides further away from a particular interaction in the network that we want to predict~\cite{abu2019mixhop}.

\begin{figure}
\centering
\includegraphics[width=0.5\textwidth]{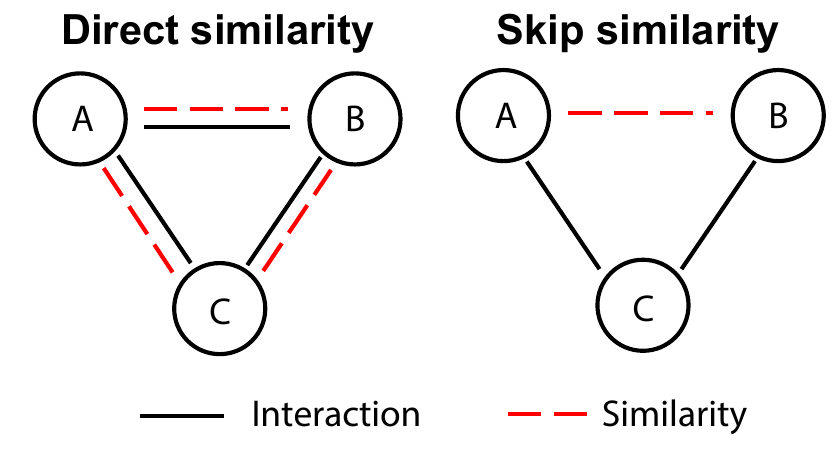}
\caption{{\bf Direct versus skip similarity.} \textit{(Left)} Traditionally, an interaction between nodes A and B implies that A and B are similar and vice versa~\cite{mcpherson2001birds}. \textit{(Right)} In contrast, in molecular interaction networks, directly interacting entities are not necessarily similar, which has been observed in numerous networks, including genetic interaction networks~\cite{costanzo2010genetic,costanzo2016global} and protein-protein interaction networks~\cite{kovacs2019network,zitnik2019evolution}. 
}
\label{fig:link}
\end{figure}

Indirect similarity between nodes that do not directly interact, \eg, the similarity in second-order interactions, has proved incredibly useful across a variety of molecular networks, including genetic interaction and protein-protein interaction networks~\cite{costanzo2010genetic,costanzo2016global,zitnik2019evolution,kovacs2019network}. This is because interactions can exist between nodes that are not necessarily similar, as illustrated in Figure~\ref{fig:link}. For example, in a drug-target interaction (DTI) network, an edge indicates that a drug binds to a target protein. Thus, two drugs are similar because they bind to the same target protein. In contrast, a drug and a target protein are not biologically similar, although they are connected by an edge in the DTI network. This example illustrates the importance of second-order interactions, which we refer to as \textit{skip similarity} (Figure~\ref{fig:link}). For this reason, we need GNNs to predict molecular interactions, not only via direct interactions but also via similarity in second-order interactions.

%%%%%

\xhdr{Present work}
Here, we present \mname, a graph neural network (GNN) method for the prediction of molecular interactions. In contrast to existing GNNs, such as GCN~\cite{kipf2017semi}, \mname specifies a neural architecture, in which neural messages are passed not only via direct interactions, referred to as direct similarity, but also via similarity in second-order interactions, referred to as \textit{skip similarity} (Figure~\ref{fig:link}). Importantly, while the principle of {\it skip similarity} governs many types of molecular interaction networks, popular GNN methods fail to capture the principle. Because of that, as we show here, they cannot fully utilize molecular interaction networks. \mname takes as input a molecular interaction network and uses it to construct a \textit{skip graph}. This second-order network representation captures the {\it skip similarity}. \mname then uses both the original graph (\ie, the input interaction network) and the skip graph to learn what is the best way to propagate and transform neural messages along edges in each graph to optimize for the discovery of new interactions. 

We evaluate \mname on four types of interaction networks, including two homogeneous networks, \ie, drug-drug interaction and protein-protein interaction networks, and two heterogeneous networks, \ie, drug-target interaction and gene-disease interaction networks. \mname outperforms baselines that use random walks, shallow network embeddings, spectral clustering, network metrics and various state-of-the-art graph neural networks~\cite{perozzi2014deepwalk,grover2016node2vec,ribeiro2017struc2vec,tang2011leveraging,kovacs2019network,abu2019mixhop,GIN}. 
%Further, the method shows a 7.9\% improvement in PR-AUC over state-of-the-art graph neural networks~\cite{kipf2017semi,kipf2016variational}. 

By examining \mname's performance in increasingly harder prediction settings when large fractions of interactions are removed from the network, we find that \mname achieves robust performance. In particular, across all interaction networks, \mname consistently outperforms all baseline methods, even when interaction networks are highly incomplete (Section~\ref{sec:exp1}-\ref{sec:missing}). We find that the robust performance of \mname can be explained by the spectral property of skip graph, as it can preserve network structure in the face of incomplete interaction information (Supplementary~\ref{sec:robustness}), which is also confirmed experimentally (Section~\ref{sec:exp4}).

Further, we examine embeddings learned by \mname and find that \mname learns biologically meaningful embeddings, whereas a regular GCN does not (Section~\ref{sec:explanation}). For example, when analyzing a drug-target interaction network, \mname generates the embedding space in which drugs are generally separated from most of proteins while still being positioned close to the proteins to which they directly bind. Lastly, in the case of the drug-drug interaction network, we use the literature search to find evidence for \mname's novel drug-drug interaction predictions~(Section~\ref{sec:exp5}). 

\xhdr{Related work} 
Existing link prediction methods belong to one of the following categories. (1) \underline{Heuristic or mechanistic methods} (\eg, \cite{lu2011link}, \cite{menche2015uncovering}, \cite{duran2018pioneering}, \cite{kovacs2019network}) calculate an index similarity score to measure the probability of a link given the network structure around the two target nodes, such as Preferential Attachment (PA)~\cite{barabasi1999emergence} and Local Path Index (LP)~\cite{lu2009similarity}. However, these methods usually make strong assumptions about the network structure and hence suffer from instability of performance~\cite{lu2011link,kovacs2019network}. (2) \underline{Direct embedding methods} generate embeddings for every node in the network capturing the node's local network topology (\eg, \cite{vzitnik2014data}, \cite{wang2018network}, \cite{xu2019network}). A popular approach is to use random walks with a skip-gram model, such as DeepWalk~\cite{perozzi2014deepwalk}, node2vec~\cite{grover2016node2vec}, and LINE~\cite{tang2015line}. The other popular approach leverages the spectral graph theory to generate a spectral embedding such as spectral clustering~\cite{tang2011leveraging}. The generated node embeddings are then fed into a decoder classifier to predict the link existing probability. (3) \underline{Neural embedding methods}, such as Graph Neural Networks (GNNs)~\cite{kipf2017semi,hamilton2017inductive}, Variational Graph Autoencoders (VGAE)~\cite{kipf2016variational,ma2018drug}, and Graph Attention Networks (GAT)~\cite{velivckovic2017graph} use neighborhood message passing scheme to generate node embeddings and these embeddings are directly optimized in an end-to-end manner by a link prediction loss (\eg, cross-entropy). {\color{black} GNNs are a powerful class of models in capturing complicated graph topology. Typically, an L-layers GNN is able to propagate information of nodes in the L-hop neighborhoods~\cite{kipf2017semi,GIN}. However, the messages of nodes farther away from the central node have discounted propagation power. Thus, the vanilla GNN is limited at capturing \textit{skip similarity}, which is from second-hop neighbors. In contrast, \mname utilizes an additional skip-graph to fully exploit this important quality for biomedical interaction network. Notably, there are recent advancements in GNN such as MixHop~\cite{abu2019mixhop}, JK-Net~\cite{xu2018representation} which are designed to capture higher order graph structures through skip connections and higher order adjacency matrix. However, they are motivated by general network model and does not propose a solution for the specific challenge of 2-hop skip similarity in biomedical network. }

In molecular interaction networks, the goal is to predict if a given pair of biomedical entities such as proteins, drugs or diseases will interact. We can divide methods for interaction prediction into three main groups. (1) \underline{Structural representation learning} generates embeddings for each entity using the entity's structural representation, such as a compound's molecular graph or a protein's amino acid sequence. The embeddings of two entities are then combined and fed into a decoder for prediction. For example, \cite{tsubaki2019compound,ozturk2018deepdta,gao2018interpretable} use graph-convolutional (GCN) and convolutional (CNN) networks on molecular graphs and gene sequence data to predict binding of drugs to target proteins. Similarly, \cite{huang2019caster,ryu2018deep,cheng2014machine} learn embedding for drugs and concatenate embeddings of drug pairs to predict drug-drug interactions. (2) \underline{Similarity-based learning} is based on the assumption that entities with similar interaction patterns are likely to interact. These methods devise a similarity measure (\eg, a graphlet-based signature of proteins in the PPI network~\cite{milenkovic2008uncovering}) and then use the measure to predict interactions based on how similar a candidate interaction is to known interactions. A variety of techniques are used to aggregate similarity values and score interactions, including matrix factorization~\cite{zhang2018predicting}, clustering~\cite{ferdousi2017computational}, and label propagation~\cite{zhang2015label}. (3) Finally, \underline{network relational learning} views the task as a network completion problem. It uses network structure together with side information about nodes to complete the network and predict interactions~\cite{zitnik2018modeling,ma2018drug,zitnik2017predicting}. \mname belongs to the structural representation learning category. 

\xhdr{Preliminaries on Graph Neural Networks (GNNs)} Next, we describe graph neural networks as they are {\color{black} one of the} state-of-the-art models for link prediction and are also the focus of our study. The input to a GNN is the network, represented by its adjacency matrix $\mathbf{A}$. Most often, the goal (output) of the GNN is to learn an embedding for each node in the network by capturing the network structure as well as node attributes. GNN can be represented as a series of neighborhood aggregations layers (\eg, \cite{kipf2017semi}):
$
    \mathbf{H}^{(l+1)} = \sigma(\widetilde{\mathbf{D}}^{-\frac{1}{2}}\widetilde{\mathbf{A}}\widetilde{\mathbf{D}}^{-\frac{1}{2}}\mathbf{H}^{(l)}\mathbf{W}) , 
$
where $\mathbf{H}^{(l)}$ is a matrix of node embeddings at the $l$-th layer, $\mathbf{H}^{(0)}$ are input node attributes, $\mathbf{W}$ is a trainable parameter matrix, $\sigma$ is a non-linear activation function, and $\widetilde{\mathbf{D}}$ and $\widetilde{\mathbf{A}}$ are the renormalized degree and adjacency matrices, defined as: $\widetilde{\mathbf{A}} = \mathbf{A} + \mathbf{I}$ and $\widetilde{\mathbf{D}}_{ii} = \sum_j \widetilde{\mathbf{A}}_{ij}$ ($\mathbf{I}$ is the identity matrix). The GNN propagates information across network neighborhoods and transforms the information in a way that is most useful for a downstream prediction tasks, such as link prediction. 

\section{Methods}

\begin{figure}[t]
\centering
\includegraphics[width = \textwidth]{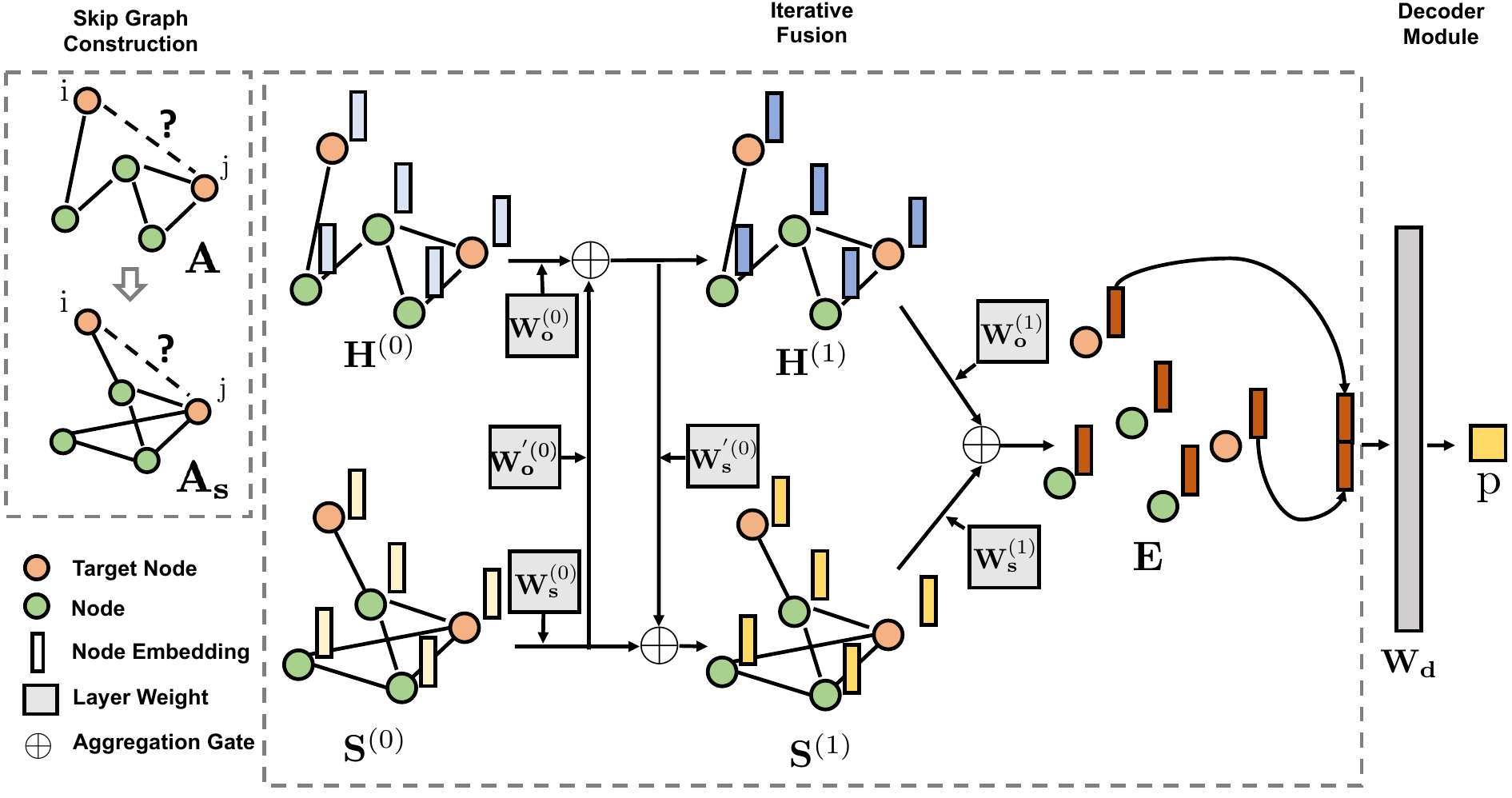} 
\caption{{\bf Neural architecture of \mname.} {\em (Left)} \mname~constructs skip graph $G_s$ (denoted by adjacency matrix $\mathbf{A}_{s}$) based on the input graph $G$ (denoted by adjacency matrix $\mathbf{A}$) using Eq.~(\ref{eq:skip2}). {\em (Middle)} Initial node embeddings, $\mathbf{H}^{(0)}$ and $\mathbf{S}^{(0)}$, are specified using side information (\eg, gene expression vectors if nodes represent genes) or generated using node2vec~\cite{grover2016node2vec}. In \mname, node embeddings are then propagated along edges of $G_s$ and $G$ and transformed through a series of computations (layers), which output powerful embeddings that can then be used for downstream prediction of interactions. Illustrated is a two-layer iterative fusion scheme. In the first layer, two GNNs with parameter weight matrices $\mathbf{W}_{o}^{(0)}$ and $\mathbf{W}_{s}^{(0)}$ (operating on $\mathbf{A}$ and $\mathbf{A}_{s}$, respectively) are fused via weight matrices $\mathbf{W}_{o}^{'(0)}$ and $\mathbf{W}_{s}^{'(0)}$ based on Eq.~(\ref{eq:fuse-gate}). This completes computations in the first layer of \mname, producing embeddings $\mathbf{H}^{(1)}$ and $\mathbf{S}^{(1)}$. In the second layer, those embeddings are transformed via $\mathbf{W}_{o}^{(1)}$ and $\mathbf{W}_{s}^{(1)}$ using Eq.~(\ref{eq:fusion-gate-final}), resulting in final embeddings $\mathbf{E}$. {\em (Right)} Embeddings $\mathbf{E}_i$ and $\mathbf{E}_j$ of target nodes $i$ and $j$ are retrieved, concatenated, and then fed into a decoder (parameterized by $\mathbf{W}_{d}$). Decoder returns ${p}_{ij}$, representing the probability that nodes ${i}$ and ${j}$ interact. }
\label{fig:iter}
\end{figure}

\begin{table}
\centering
\caption{\label{tb:notation}{\bf Notation used in \mname.}}

\begin{tabular}{ll}

\toprule Notation & Definition \\
  \midrule 
  $G:\{\mathcal{V}, \mathcal{E}\}$ & Graph  with nodes $\mathcal{V}$ and edges $\mathcal{E}$ \\
  $\mathbf{D}, \mathbf{A} \in {\color{black} \mathbb{N}}^{N \times N}$ & Degree and adjacency matrices for graph $G$ \\
  $\widetilde{\mathbf{D}}, \widetilde{\mathbf{A}} \in {\color{black} \mathbb{N}}^{N \times N}$ & Normalized degree and adjacency matrices for $G$ \\
  $\mathbf{X} \in \mathbb{R}^{N \times D}$ & $D$-dimensional node embeddings \\
  ${e}_{ij} \in \{0,1\}$ & Ground-truth interaction between nodes $i$ and $j$\\
  \midrule
  $G_s$ & Skip graph \\
  $\mathbf{D}_{s}, \mathbf{A}_{s} \in {\color{black} \mathbb{N}}^{N \times N}$ & Degree and adjacency matrices for  $G_s$ \\
  $\widetilde{\mathbf{D}_{s}}, \widetilde{\mathbf{A}_{s}} \in \mathbb{R}^{N \times N}$ & Normalized degree and adjacency matrices for  $G_s$ \\ \midrule
  $\mathbf{H}^{(l)}, \mathbf{S}^{(l)}$ & Node embeddings for $G$ and $G_s$, in layer $l$\\
  $\mathbf{E}$ & Final node embeddings \\
  ${p}_{ij} \in [0, 1]$ & Probability of interaction between nodes ${i}$ and ${j}$\\
  ${y}_{ij} \in \{0, 1\}$ & Binary indicator of interaction between nodes ${i}$ and ${j}$\\
  $\mathcal{L} \in \mathbb{R}$ & {\color{black} Binary cross-entropy loss} \\ \midrule
  $\mathbf{W}_{o}^{(l)}, \mathbf{W}_{s}^{(l)}$ & Weight matrix for original ($o$) and skip ($s$) graphs, layer $l$ \\
  $\mathbf{W}_{o}^{'(l)}$ & Weight matrix for skip-to-original-graph fusion\\ 
  $\mathbf{W}_{s}^{'(l)}$ & Weight matrix for original-to-skip-graph fusion\\ 
  $\mathbf{W}_{d}, {b}$ & Decoder weight matrix and bias parameter\\ \bottomrule
\end{tabular}
\end{table}

\mname is a graph neural network uniquely suited for molecular interactions. \mname takes as input a molecular interaction network and uses it to construct a skip graph, which is a second-order network representation capturing the \textit{skip similarity}. \mname then specifies a novel graph neural network architecture that fuses the original and the skip graph to accurately and robustly predict new molecular interactions. Notations are described in Table~\ref{tb:notation}.

\xhdr{Problem formulation}
Consider an interaction network $G$ on $N$ nodes representing biomedical entities $\mathcal{V}$  (\eg, drugs, proteins, or diseases) and $M$ edges $\mathcal{E}$ representing interactions between the entities. For example, $G$ can be a drug-target interaction network recording information on how drugs bind to their protein targets~\cite{luo2017network}. For every pair of entities $i$ and $j$, we denote their interaction with a binary indicator ${e}_{ij} \in \{0,1\}$, indicating the experimental evidence that $i$ and $j$ interact (\ie, ${e}_{ij}=1$) or the absence of evidence for interaction (\ie, ${e}_{ij}=0$). We denote the adjacency matrix of $G$ as $\mathbf{A}$, where $\mathbf{A}_{ij}$ is 1 if nodes $i$ and $j$ are connected (${e}_{ij}=1$) in the graph and otherwise 0 (${e}_{ij}=0$). Further, $\mathbf{D}$ is the degree matrix, a diagonal matrix, where $\mathbf{D}_{ii}$ is the degree of node $i$.
\begin{problem*}[Molecular Interaction Prediction]
Given a molecular interaction network $G=(\mathcal{V}, \mathcal{E})$, we aim to learn a mapping function $f: \mathcal{E} \to [0, 1]$ from edges to probabilities such that $f({i,j})$ optimizes the probability that nodes $i$ and $j$ interact. 
\end{problem*}

\subsection{Construction of the skip graph}\label{sec:skip-graph}

Next, we describe skip graphs, the key novel representation of interaction networks that allow for effective use of GNNs for predicting interactions.  We realize \textit{Skip similarity} by encouraging the GNN model to embed skipped nodes close together in the embedding space. To do that, we construct skip graph $G_s$, in two-hop neighbors are connected by edges. This construction creates paths in $G_s$ along which neural messages can be exchanged between the skipped nodes. 

Formally, we use the following operator to obtain the skip graph's adjacency matrix $\mathbf{A}_{{s}}$:
\begin{equation*}\label{eq:skip}    
\mathbf{A}_{{s}}^{ij} = \left\{ 
                \begin{array}{ll}
                  1~~\mathrm{if}~\exists~k~\mathrm{s.t.}~{(i,k)}~\in~\mathcal{E}~\mathrm{and}~{(k,j)}~\in~\mathcal{E}\\
                  0~~\mathrm{otherwise.}
                \end{array}
            \right. 
\end{equation*}
The corresponding degree matrix is $\mathbf{D}_{{s}}^{ii} = \sum_{j} \mathbf{A}_{{s}}^{ij}.$
An efficient way to implement the skip graph is through matrix multiplication:
\begin{equation}\label{eq:skip2}
    \mathbf{A}_{{s}} = \mathrm{sign}(\mathbf{A}\mathbf{A}^\mathrm{T}),
\end{equation}
where: $\mathrm{sign}(x)$ is the sign function, $\mathrm{sign}(x) = 1$ if $x > 0$ and $0$ otherwise, which is applied element-wise on $\mathbf{A}\mathbf{A}^\mathrm{T}$. It counts the number of two-hop paths from node $\mathrm{i}$ to $\mathrm{j}$. Hence, if an entry for node $\mathrm{i}, \mathrm{j}$ in $\mathbf{A}\mathbf{A}^\mathrm{T}$ is larger than 0, it means there exists a skipped node between node $i, j$. Then, we convert the positive entry into 1 to construct the skip graph's adjacent matrix. Given this skip graph, we proceed to describe the full \mname model.

\subsection{The \mname~model}
In this section, we describe how we leverage the skip graph for link prediction. After we generate the novel skip graph from Section~\ref{sec:skip-graph}, we propose an iterative fusion scheme for \mname~to allow the skip graph and the original graph to learn from each other for better integration. Lastly, a decoder is used to output a probability measuring if the given pair of molecular entities interact. 

\subsubsection{Iterative fusion}\label{sec:iterative-fusion}

We want a model to automatically learn how to balance between \textit{direct similarity} and \textit{skip similarity} in the final embedding. We design an iterative fusion scheme with aggregation gates to combine both similarity information. The motivation is that to represent biomedical entity to its fullest extent, node embedding must capture its complicated bioactive functions with \textit{skip/direct similarities}. Hence, instead of simply concatenating the output node embeddings from the GNN output of the original graph $G$ that captures \textit{direct similarity} and skip graph $G_s$ that captures \textit{skip similarity}, we allow two GNNs on $G$ and $G_s$ to interact with each other iteratively via the following propagation rules (see Figure~\ref{fig:iter}):
\begin{equation}
\label{eq:fuse-gate}
\mathbf{H}^{(l+1)} = \sigma (\mathrm{AGG}(\mathbf{F}\mathbf{H}^{(l)}\mathbf{W}_{o}^{(l)},\mathbf{F}_{{s}} \mathbf{S}^{(l)}\mathbf{W}_{o}^{'{(l)}}))\text{,  } \hspace{3mm}  \mathbf{S}^{(l+1)}  = \sigma (\mathrm{AGG}(\mathbf{F}_{{s}}\mathbf{S}^{(l)}\mathbf{W}_{{s}}^{(l)},\mathbf{F}\mathbf{H}^{(l+1)}\mathbf{W}_{{s}}^{'(l)})),    \\
\end{equation}
where $\mathbf{F} = \widetilde{\mathbf{D}}^{-\frac{1}{2}}\widetilde{\mathbf{A}}\widetilde{\mathbf{D}}^{-\frac{1}{2}} \text{,  } \hspace{3mm} \mathbf{F}_{{s}} = \widetilde{\mathbf{D}}_{\mathrm{s}}^{-\frac{1}{2}}\widetilde{\mathbf{A}}_{{s}}\widetilde{\mathbf{D}}_{{s}}^{-\frac{1}{2}}. $ Here, $\mathbf{H}^{(l)}, \mathbf{S}^{(l)}$ are node embeddings at the $l$-th layer from direct similarity graph $G$ and \textit{skip similarity} graph $G_S$, respectively.  $\mathbf{F}, \mathbf{F}_{{s}}$ are the re-normalized adjacency matrices from direct similarity and \textit{skip similarity}, respectively. And $\mathbf{W}_{o}^{(l)}, \mathbf{W}_{o}^{'(l)}, \mathbf{W}_{{s}}^{(l)}, \mathbf{W}_{{s}}^{'(l)}$ are the transformed weights for layer $l$. $\mathbf{H}^{(0)}$ and $\mathbf{S}^{(0)}$ are set to be $\mathbf{X}$, the input node attributes generated from node2vec. The aggregate gate $\mathrm{AGG}$ in Eq.~(\ref{eq:fuse-gate}) can be a summation, a Hadamard product, max-pooling, or some other aggregation operator~\cite{cao2020comprehensive}. Empirically, we find that summation gate has the best performance. $\sigma()$ is the activation function and we use $\mathrm{ReLU}(\cdot) = \max(\cdot, 0)$ to add non-linearity in the propagation. 

In each iteration, the node embedding for original graph $\mathbf{H}^{(l+1)}$ is first updated with its previous layer's node embedding $\mathbf{H}^{(l)}$, combined with  skip graph embedding $\mathbf{S}^{(l)}$. After obtaining the updated original graph embedding $\mathbf{H}^{(l+1)}$, we then update the skip graph embedding $\mathbf{S}^{(l+1)}$ in a similar fashion. 

This update rule is very different from simple concatenation as it is an iterative process where each update of the node embedding for each graph is affected by the most \textit{recent} node embedding from both graphs. This way, two embedding are learned to find the best dependency structure between each other and fuse into one final embedding instead of a simple concatenation. In the last layer, final node embedding $\mathbf{E}$ is obtained through:
\begin{equation}\label{eq:fusion-gate-final}
       \mathbf{E}  = \mathrm{AGG}(\mathbf{F}\mathbf{H}^{({\color{black} 1})}\mathbf{W}_{o}^{({\color{black} 1})},~\mathbf{F}_{{s}} \mathbf{S}^{({\color{black} 1})}\mathbf{W}_{s}^{({\color{black} 1})}),
\end{equation}
where {\color{black} $(1)$ is the index for the last layer} and {\color{black} $\mathrm{AGG}$ is the summation gate}. As in the motivation, we are interested only in up to second order neighbors, thus we use two layers GNN, see Figure~\ref{fig:iter}. We don't use activation function here as it does not require an extra non-linear transformation to be fed into the decoder network. Empirically, we show this fusion scheme boosts predictive performance in Section~\ref{sec:exp4}. 

\subsubsection{\mname decoder}\label{sec:decoder}

Given the target nodes $({i}, {j})$ and their corresponding node embedding $\mathbf{E}_{i}, \mathbf{E}_{j}$, we implement a neural network as a decoder to first combine $\mathbf{E}_{i}, \mathbf{E}_{j}$ to obtain an input embedding through a $\mathrm{COMB}$ function (e.g., concatenation, sum, Hadamard product). Then, the combined embedding is fed into a neural network parametrized by weight $\mathbf{W}_{d}$ and bias ${b}$ as a binary classifier to obtain probability ${p}_{ij}$:
\begin{equation}\label{eq:decoder}
    {p}_{ij} = \sigma(\mathbf{W}_{d} \mathrm{COMB}(\mathbf{E}_{i}, \mathbf{E}_{j}) + {b}),   
\end{equation}
where ${p}_{ij}$ represents the probability that nodes ${i}$ and ${j}$ interact (\ie, $f({i},{j})$. We use concatenation as the $\mathrm{COMB}$ function as it consistently yield the best performance across different types of networks. 

\begin{algorithm}[t]
\textbf{Input:} interaction network $G$ with adjacent matrix $\mathbf{A}$\\
\textbf{Node Embedding Generation, \eg:} 
\\$\mathbf{X} \leftarrow \mathrm{node2vec}(\mathbf{A})$ \\
\textbf{Skip Graph Construction} (Section~\ref{sec:skip-graph}): \\ $\mathbf{A}_\mathrm{s} \leftarrow \mathrm{SkipGraph}(\mathbf{A})$ via Eq.~ (\ref{eq:skip2})\\
\For{$t = 1 \ldots T_{\max}$}{
    sample mini-batch of training node pairs $\mathcal{M} \subseteq \mathcal{E}$ with corresponding labels $\mathrm{y}$ \\
    {\color{black}
    \textbf{Iterative Fusion} (Section~\ref{sec:iterative-fusion}): \\
    $\mathbf{H}^{(1)} \leftarrow \mathrm{FusionGate}(\mathbf{H}^{(0)})$ via Eq.~ (\ref{eq:fuse-gate}) \\
    $\mathbf{S}^{(1)} \leftarrow \mathrm{FusionGate}(\mathbf{S}^{(0)})$ via Eq.~ (\ref{eq:fuse-gate}) \\
    }
    $\mathbf{E} \leftarrow \mathrm{FuseGate}(\mathbf{H}^{({\color{black} 1})}, \mathbf{S}^{({\color{black} 1})} )$ via Eq.~ (\ref{eq:fusion-gate-final})\\
    \textbf{Decoder} (Section~\ref{sec:decoder}):\\
    ${p}_{ij} \leftarrow \textbf{decode}(\mathbf{E})$ via Eq.~ (\ref{eq:decoder})\\
    Compute the loss value $\mathcal{L}$ using ${p}_{ij}$ and ${y}$ (Section~\ref{sec:algorithm}) and update model parameters via gradient descent
}
\caption{\label{algo} The \mname~Algorithm}
\end{algorithm}

\subsection{The \mname algorithm}\label{sec:algorithm}

The overall algorithm is shown in Algorithm~\ref{algo}. 
Here, we only leverage accessible network information (adjacent matrix $\mathbf{A}$ of the network $G$) to predict links. In all experiments, we initialize embeddings using node2vec~\cite{grover2016node2vec} as: $\mathbf{X} = \mathrm{node2vec}(\mathbf{A}).$ 

Second, we construct the skip graph with adjacent matrix $\mathbf{A}_{s}$ via Eq.~(\ref{eq:skip2}) to capture the \textit{skip-similarity} principle. 
Next, at every step, a mini-batch of interaction pairs $\mathcal{M}$ with labels ${y}$ is sampled. Then, two graph convolutions networks are used for the original graph and the skip graph respectively. In the propagation step, we use iterative fusion~(Eq.~(\ref{eq:fuse-gate})) to naturally combine embeddings convolved on the original graph and on the skip graph, corresponding to \textit{direct} and \textit{skip similarity}, respectively. In the last layer, embeddings are stored in $\mathbf{E}$. We then retrieve the embeddings for each node in the mini-batched pairs $\mathcal{M}$ and concatenate them to feed into decoder~(Eq.~(\ref{eq:decoder})). 

During training, we optimize the \mname's parameters $\mathbf{W}_{o}^{(l)}$, $\mathbf{W}_{{o}}^{'(l)}, \mathbf{W}_{{s}}^{(l)}$, $\mathbf{W}_{{s}}^{'(l)}$, $\mathbf{W}_{d}$, ${b}$ in an end-to-end manner through a {\color{black} binary cross-entropy loss}: 
$
    \mathcal{L} = \sum_{{(i,j)} \in \mathcal{M}} {y}_{ij}~\mathrm{log}~{p}_{ij} + (1 -  {y}_{ij}) ~\mathrm{log}~({1-{p}_{ij}}),
$
where ${y}_{ij}$ is the true label for nodes ${i}$ and ${j}$ that are sampled during training via mini-batching,  ${(i,j)} \in \mathcal{M}$, and $\mathcal{M}$ is a mini-batch of interaction pairs.
After the model is trained, it can be used to make predictions. Given two entities $i$ and $j$, the model predicts probability $f(i,j)$ that $i$ and $j$ interact. 

\section{Results}
We conduct a variety of experiments to investigate the predictive power of \mname (Section~\ref{sec:exp1}). We then study the method's robustness to noise and missing data (Section~\ref{sec:missing}) and demonstrate
the skip similarity principle (Section~\ref{sec:explanation}). Next, we conduct ablation studies to examine contributions of each of \mname's components towards the final \mname performance (Section~\ref{sec:exp4}). Finally, we investigate novel predictions made by \mname (Section~\ref{sec:exp5}).

\subsection{Data and experimental setup}

\begin{table}[t]
\centering
\caption{\label{tab:data}{\bf Data statistics.} `A' indicates average node degree. } 
\begin{tabular}{l|lccc}
\toprule 
Dataset & Prediction task & $\#$ nodes & $\#$ edges & A \\\midrule
DTI & Drug-target interaction & 7,343 & 15,139 & 4.12 \\
DDI & Drug-drug interaction & 1,514 & 48,514 & 64.09 \\
PPI& Protein-protein interaction & 5,604 & 23,322 & 8.32 \\
GDI & Gene-disease interaction & 19,783 & 81,746 & 8.26 \\
\bottomrule
\end{tabular} 
\end{table}

Next we provide details on molecular interaction datasets, baseline methods, and experimental setup. 

\subsubsection{Molecular interaction networks} 
We consider four publicly-available network datasets. (1) \textit{BIOSNAP-DTI}~\cite{biosnapnets} contains 5,018 drugs that target 2,325 protein through 15,139 drug-target (DTI) interactions. (2) \textit{BIOSNAP-DDI}~\cite{biosnapnets} consists of 48,514 drug-drug interactions (DDIs) between 1,514 drugs extracted from drug labels and biomedical literature. (3) \textit{HuRI-PPI}~\cite{luck2019reference} is the human reference protein-protein interaction network generated by multiple orthogonal the high-throughput yeast two-hybrid screens. We use HI-III network, which has 5,604 proteins and 23,322 interactions. (4) Finally, we consider \textit{DisGeNET-GDI}~\cite{10.1093/nar/gkz1021} collects curated gene-disease interactions (GDIs) from GWAS studies, animal models and scientific literature. The dataset has 81,746 interactions between 9,413 genes and 10,370 diseases. Dataset statistics are described in Table~\ref{tab:data}.

\subsubsection{\mname implementation and hyperparameters} 

We implemented \mname~using PyTorch deep learning framework\footnote{The source code implementation of \mname is available at \url{https://github.com/kexinhuang12345/SkipGNN}.}. We use a server with 2 Intel Xeon E5-2670v2 2.5GHZ CPUs, 128GB RAM and 1 NVIDIA Tesla P40 GPU. We set optimization parameters as follows: learning rate is 5e-4 using the Adam optimizer~\cite{kingma2014adam}, mini-batch size is $|\mathcal{M}| = 256$, epoch size is 15, and dropout rate is 0.1. We set hyper-parameters using 10 runs random search based on best average prediction performance on validation set of DTI task. We find the setup is robust in other datasets. {\color{black} The ranges of hyper-parameters are set as follows: learning rate: [1e-3, 5e-4, 1e-4, 5e-5]; mini-batch size [32, 64, 128, 256, 512]; dropout rate [0, 0.05, 0.1, 0.2]; hidden size [16, 32, 64, 128].} Specifically, we set hidden size in the first layer as $d^{(1)}=64$ and hidden size in the second layer as $d^{(2)}=16$. 
\subsubsection{Baseline methods}

We compare \mname to seven powerful predictors of molecular interactions from network science and graph machine-learning fields. From machine learning, we use three direct network embedding methods: \textbf{DeepWalk}~\cite{perozzi2014deepwalk}, \textbf{node2vec}~\cite{grover2016node2vec}, and we also include \textbf{struc2vec}~\cite{ribeiro2017struc2vec}. The latter method is conceptually distinct by leveraging local network structural information, while the former methods use random walks to learn embeddings for nodes in the network. {\color{black} Further, we examine five graph neural networks: \textbf{VGAE}~\cite{kipf2016variational}, \textbf{GCN}~\cite{kipf2017semi}, \textbf{GIN}~\cite{GIN}, \textbf{JK-Net}~\cite{xu2018representation} and \textbf{MixHop}~\cite{abu2019mixhop}.} They all use the same input encoding as \mname. From network science, we consider \textbf{Spectral Clustering}~\cite{tang2011leveraging}. We also use \textbf{L3}~\cite{kovacs2019network} heuristic, which was recently shown to outperform over 20 network science methods for the problem of PPI prediction. Further details on baseline methods, their implementation and parameter selection are in supplementary.

\begin{table}[t]
\color{black}
\centering
\caption{\label{tab:q1}{\bf Predictive performance.} \mname~achieves the best performance across all metrics and tasks compared to baselines. Results of five independent runs on DDI, PPI, DTI and GDI tasks on state of the art link prediction algorithms.} 

\centering
\adjustbox{max width=0.45\textwidth}{
\begin{tabular}{l|lccc}
\toprule     
Task & Method & PR-AUC & ROC-AUC & Rank\\ \midrule
    \multirow{11}{*}{DTI}
    & DeepWalk &$0.753 \pm 0.008$ &$0.735 \pm 0.009$ &  10 \\
    & node2vec &$0.771 \pm 0.005$ &$0.720 \pm 0.010$ & 9 \\
    & struc2vec &$0.677 \pm 0.007$ &$0.656 \pm 0.010$ & 11 \\
    \cline{2-5}
    & SC &$0.818 \pm 0.007$ & $0.743 \pm 0.008$ & 8\\
    & L3 &$0.891 \pm 0.004$ &$0.793 \pm 0.006$ & 6\\
    \cline{2-5}
    & VGAE & $0.853 \pm 0.010$ & $0.800 \pm 0.010$ & 7 \\
    & GCN  &$0.904 \pm 0.011$ &$0.899 \pm 0.010$ & 5 \\
    & GIN  &$0.922 \pm 0.004$ &$0.907 \pm 0.006$ & 3 \\
    & JK-Net  &$0.921 \pm 0.006$ &$0.907 \pm 0.008$ & 4\\
    & MixHop  &$0.921 \pm 0.006$ &$0.920 \pm 0.004$ & 2 \\
    \cline{2-5}
    & \mname~& $\bf{0.928 \pm 0.006}$ & $\bf{0.922 \pm 0.004}$ & 1 \\
    \midrule
    \multirow{11}{*}{DDI}
    & DeepWalk &$0.698 \pm 0.012$ &$0.712 \pm 0.009$ & 10 \\
    & node2vec &$0.801 \pm 0.004$ & $0.809 \pm 0.002 $ & 8 \\
    & struc2vec &$0.643 \pm 0.012$ &$0.654 \pm 0.007$ & 11 \\
    \cline{2-5}
    & SC & $0.749 \pm 0.009$ &$0.816 \pm 0.006$ & 9\\
    & L3 &$0.860 \pm 0.004$ &$0.869 \pm 0.003$ & 4\\
    \cline{2-5}
    & VGAE  &$0.844 \pm 0.076$ & $0.878 \pm 0.008$ & 7 \\
    & GCN & $0.856 \pm 0.005$ &$0.875 \pm 0.004$ & 5\\
    & GIN  &$0.856 \pm 0.005$ &$0.876 \pm 0.003$ & 5 \\
    & JK-Net  &$\bf{0.870 \pm 0.009}$ &$0.885 \pm 0.005$ & 1\\
    & MixHop  &$0.861 \pm 0.006$ &$0.879 \pm 0.004$ &3 \\
    \cline{2-5}
    & \mname~& $0.866 \pm 0.006$ &$\bf{0.886 \pm 0.003}$ & 2 \\ 
    \bottomrule
    \end{tabular}
    }
    \quad
    \adjustbox{max width=0.45\textwidth}{%
    \begin{tabular}{l|lccc}
    \toprule     
    Task & Method & PR-AUC & ROC-AUC & Rank \\ \midrule    
    \multirow{11}{*}{PPI}
    & DeepWalk &$0.715 \pm 0.008 $ &$0.706 \pm 0.005$ & 11 \\
    & node2vec &$0.773 \pm 0.010$ &$0.766 \pm 0.005$ & 10 \\
    & struc2vec &$0.875 \pm 0.004$ &$0.868 \pm 0.006$ & 8 \\
    \cline{2-5}
    & SC  & $0.897 \pm 0.003$& $0.859 \pm 0.003$ & 7\\
    & L3 &$0.899 \pm 0.003$ &$0.861 \pm 0.003$ & 6 \\
    \cline{2-5}
    & VGAE  & $0.875 \pm 0.004$ &$0.844 \pm 0.006$ & 8\\
    & GCN  &$0.909 \pm 0.002$ &$0.907 \pm 0.006$ & 4\\
    & GIN  &$0.907 \pm 0.004$ &$0.897 \pm 0.006$ & 5\\
    & JK-Net  &$0.912 \pm 0.003$ &$0.901 \pm 0.005$ & 3 \\
    & MixHop  &$0.909 \pm 0.004$ &$0.913 \pm 0.003$ & 2 \\
    \cline{2-5}
    & \mname~& $\bf{0.921 \pm 0.003}$ &$\bf{0.917 \pm 0.004}$ & 1 \\ 
    \midrule
    \multirow{11}{*}{GDI}
    & DeepWalk &$0.827 \pm 0.007 $ &$0.832 \pm 0.003$ & 11\\
    & node2vec &$0.828 \pm 0.006$ &$0.834 \pm 0.003$ & 10  \\
    & struc2vec &$0.910 \pm 0.006$ &$0.909 \pm 0.005$ & 4 \\
    \cline{2-5}
    & SC  & $0.905 \pm 0.002$& $0.863 \pm 0.003$ & 6 \\
    & L3 &$0.899 \pm 0.001$ &$0.832 \pm 0.001$ & 8\\
    \cline{2-5}
    & VGAE  & $0.902 \pm 0.006$ &$0.873 \pm 0.009$ & 7\\
    & GCN  &$0.909 \pm 0.002$ &$0.906 \pm 0.006$ & 5\\
    & GIN  &$\bf{0.916 \pm 0.004}$ &$0.900 \pm 0.005$ & 1 \\
    & JK-Net  &$0.891 \pm 0.049$ &$0.898 \pm 0.002$ & 9 \\
    & MixHop  &$0.912 \pm 0.005$ &$\bf{0.916 \pm 0.004}$ & 3\\
    \cline{2-5}
    & \mname~& $0.915 \pm 0.003$ &$0.912 \pm 0.004$ & 2 \\ 
    \bottomrule
\end{tabular}
}
\end{table}

\begin{table}[t]
\color{black}
\caption{\label{tab:overall_perf}{\color{black} {\bf Predictive performance ranking and statistical testing.} We rank each tested method's PR-AUC in each dataset and computes the average rank and also computes the performance difference from SkipGNN using Wilcoxon signed-rank test. \mname~has the highest rank compared with 10 other baselines and its performance gain is statistically significant.}} 
\centering
    \adjustbox{max width=0.98\textwidth}{%
    \begin{tabular}{l|ccc|cc|ccccc|c}
    \toprule
    Method & DeepWalk & node2vec & struc2vec & SC & L3 & VGAE & GCN & GIN & JK-Net & MixHop & SkipGNN \\ \midrule
    Average Rank & 10.5 & 9.25 & 8.50 & 7.50 & 6.00 & 7.25 & 4.75 & 3.75 & 4.00 & 2.50  & 1.50 \\
    p-value & <.001 & <.001 & .006 & .003 & .016 & .005 & .012 & .017 & .025 & .042 & N/A \\
    \bottomrule
    \end{tabular}
    }
\end{table}

\subsubsection{Experimental setup}

In all our experiments, we follow an established evaluation strategy for link prediction~(\eg, \cite{zhang2018link,zitnik2018modeling}). 
We divide each dataset into train, validation, and test sets in a 7:1:2 ratio, which yields positive examples (molecular interactions). {\color{black} We generate negative counterparts by sampling the complement set of positive examples. The cardinality of negative samples are set to be the same as positive data points.} For every experiment, we conduct five independent runs with different random splits of the dataset. We select the best performing model based on the loss value on the validation set. The performance of selected model is calculated on the test set. To calculate prediction performance, we use: (1) area under precision-recall curve~(PR-AUC): $\text{PR-AUC} = \sum_{k = 1}^{n} \mathrm{Prec}(k) \Delta \mathrm{Rec}(k),$ 
    where $k$ is $k$-th precision/recall operating point ($\mathrm{Prec}(k), \mathrm{Rec}(k)$); and (2) area under the receiver operating characteristics curve~(ROC-AUC): $\text{ROC-AUC} = \sum_{k = 1}^{n} \mathrm{TP}(k) \Delta \mathrm{FP}(k),$ where $k$ is $k$-th true-positive and false-positive operating point ($\mathrm{TP}(k), \mathrm{FP}(k)$).
Higher values of PR-AUC and ROC-AUC indicate better predictive performance. {\color{black} In addition to the PR-AUC and ROC-AUC, we rank each method in each dataset based on its PR-AUC and provide the average rank of a method across four datasets. The rank suggests the overall performance of the method compared to others. To further show the performance gain of \mname, we resort to statistical test. For each method, we take the ROC-AUC and PR-AUC of each run for each dataset as the data samples. Then, we compute the p-value for Wilcoxon signed-rank test between SkipGNN and the compared method. }

\subsection{Predicting molecular interactions} \label{sec:exp1}

We start by evaluating \mname on four distinct types of molecular interactions, including drug-target interactions, drug-drug interactions, protein-protein interactions, and gene-disease interactions, and we then compare \mname's performance to baseline methods.

In each interaction network, we randomly mask 30\% interactions as the holdout validation (20\%) and test (10\%) sets. The remaining 70\% interactions are used to train the \mname and each of the baselines. After training, each method is asked to predict whether pairs of entities in the test set will likely interact.

{\color{black} We report results in Table~\ref{tab:q1} and the method rank, along with the p-values for statistical test are provided in Table~\ref{tab:overall_perf}.  We see that \mname~is the top performing method out of 11 methods across all molecular interaction networks. SkipGNN has the best predictive performance for DTI and PPI datasets and has the second best performance in DDI and GDI datasets, with an average rank of 1.5. In contrast, the best performing baseline MixHop has average rank of 2.5, as it sometimes is worse than JK-Net and GIN. We also see that SkipGNN’s improvement over all baselines is statistically significant (<.05). To show the usefulness of skip graph, we compare with GCN-backend baselines GCN and VGAE. We see up to 2.7\% improvement of \mname~over GCN and up to 8.8\% improvement over VGAE on PR-AUC. Since GCN and VGAE can only use \textit{direct similarity}, this finding provides evidence that considering \textit{skip similarity} and \textit{direct similarity} together, as is made possible by \mname, is important to be able to accurately predict a variety of molecular interactions. Compared to direct embedding methods, \mname has up to 28.8\% increase over DeepWalk, 20.4\% increase over node2vec, and 15.6\% over spectral clustering on PR-AUC. These results support previous observations~\cite{zitnik2018modeling} that graph neural networks can learn more powerful network representations than direct embedding methods. Finally, all baselines vary in performance across datasets/tasks while \mname consistently yields the most powerful predictor.}

\subsection{Robust learning on incomplete interaction networks}\label{sec:missing}

\begin{figure}
    \centering
    \includegraphics[width=0.8\textwidth]{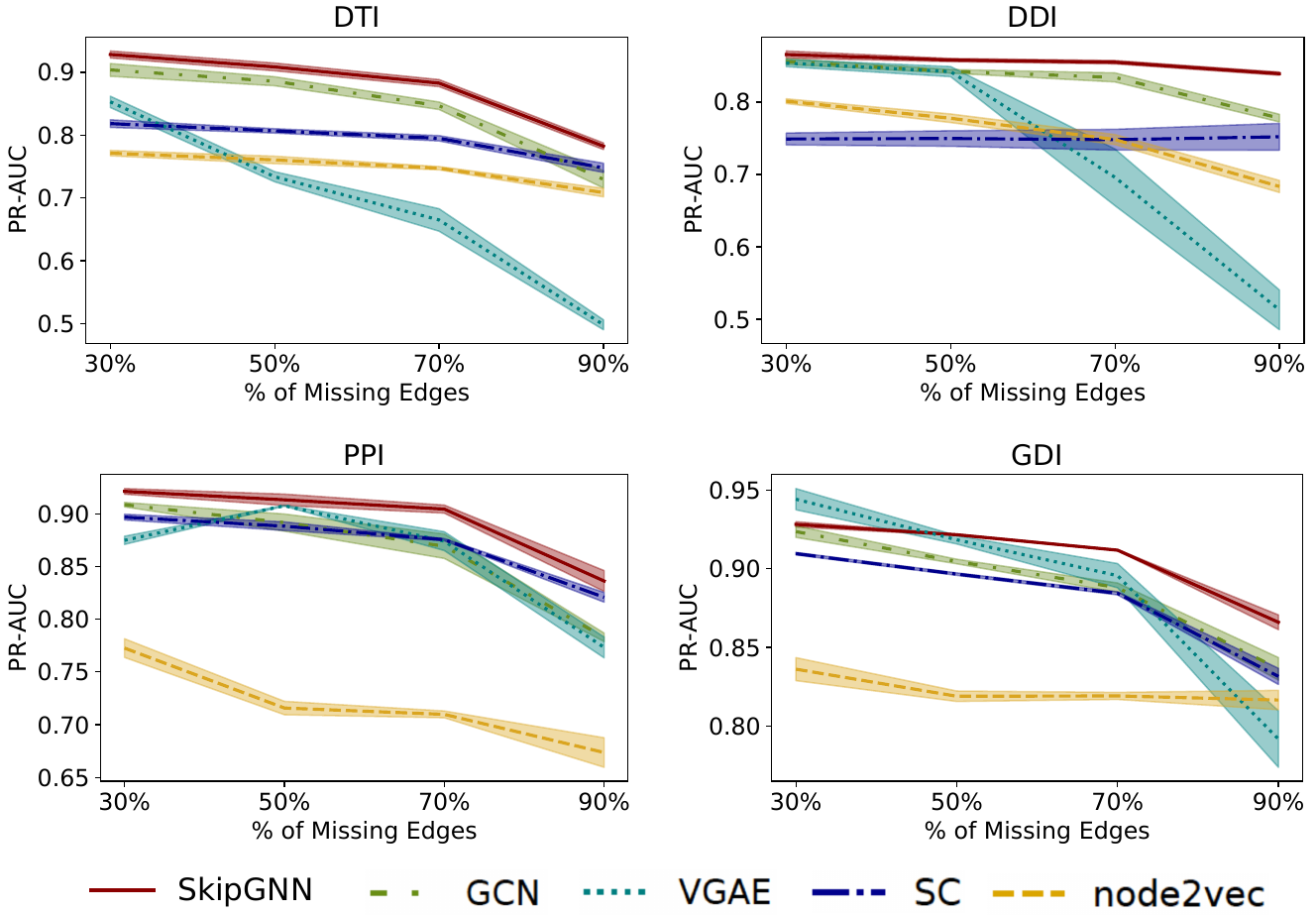}
    \caption{{\bf Predictive performance as a function of network incompleteness.} \mname~provides robust result in varying fraction of missing edges. Five-fold average with 95\% confidence interval for PR-AUC against various fractions of missing edges on four prediction tasks: drug-target interaction prediction (DTI), drug-drug interaction prediction (DDI), protein-protein interaction prediction (PPI) and gene-disease interaction prediction (GDI) on node2vec, Spectral Clustering (SC), Variational Graph Auto-Encoder (VGAE), Graph Convolutional Network (GCN), and \mname. We omit DeepWalk as it has similar performance as node2vec. SkipGNN consistently shows the best performance even when networks are highly incomplete.}
\label{fig:missing}
\end{figure}

Next, we test \mname's performance on incomplete interaction networks. Due to knowledge gaps in biology, many of today's interaction networks are incomplete and thus it is crucial that methods are robust and able to perform well even when many interactions are missing. 

In this experiment, we let each method be trained on 10\%, 30\%, 50\%, and 70\% of edges in the DTI, DDI, and PPI datasets and predict on the rest of the data (we use 10\% of test edges as validation set for early stopping). 

Results in Figure~\ref{fig:missing} show that \mname~gives the most robust results among all the methods. In all tasks, \mname achieves strong performance even when having access to only 10\% of the interactions. Further, in almost every percentage point, \mname~is better than the baselines. In addition, we see that VGAE is not robust as its performance dropped to around 0.5 PR-AUC in highly-incomplete settings on DTI and DDI tasks. Performance of node2vec and GCN steadily improve as the percentage of seen edges increases. Further, while spectral clustering is robust to incomplete data, its performance varies substantially with tasks. We conclude that \mname~is robust and is especially appropriate for data-scarce networks. 

\subsection{\mname learns meaningful embedding spaces} \label{sec:explanation}
\begin{figure}
\centering
\includegraphics[width=\textwidth]{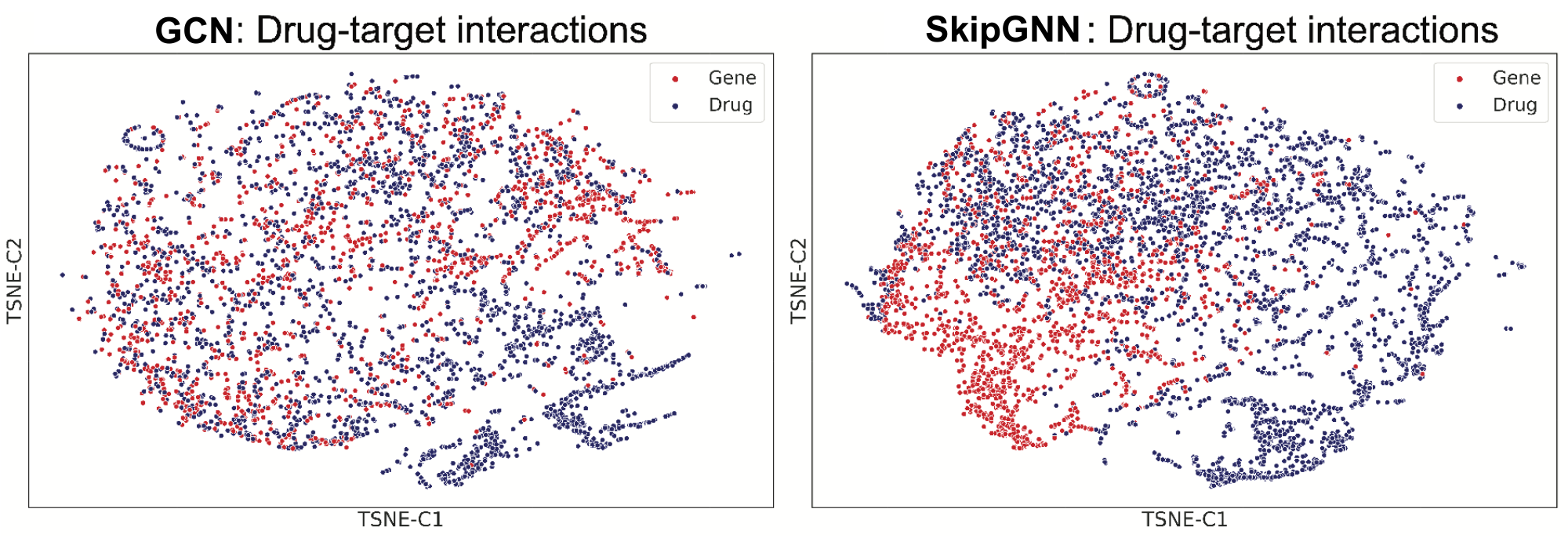}
\caption{{\bf Visualizations of drug-target interaction network.} GCN does not distinguish drug and target gene as it only captures direct similarity whereas \mname is able to distinct drug and target gene embeddings, confirming its ability to capture \textit{skip similarity}. We use GCN and \mname on the drug-target interaction dataset to learn drug/target embeddings, which are visualized using t-SNE.}
\label{fig:explain-dti}
\end{figure}

\begin{figure}
\centering
\includegraphics[width=\textwidth]{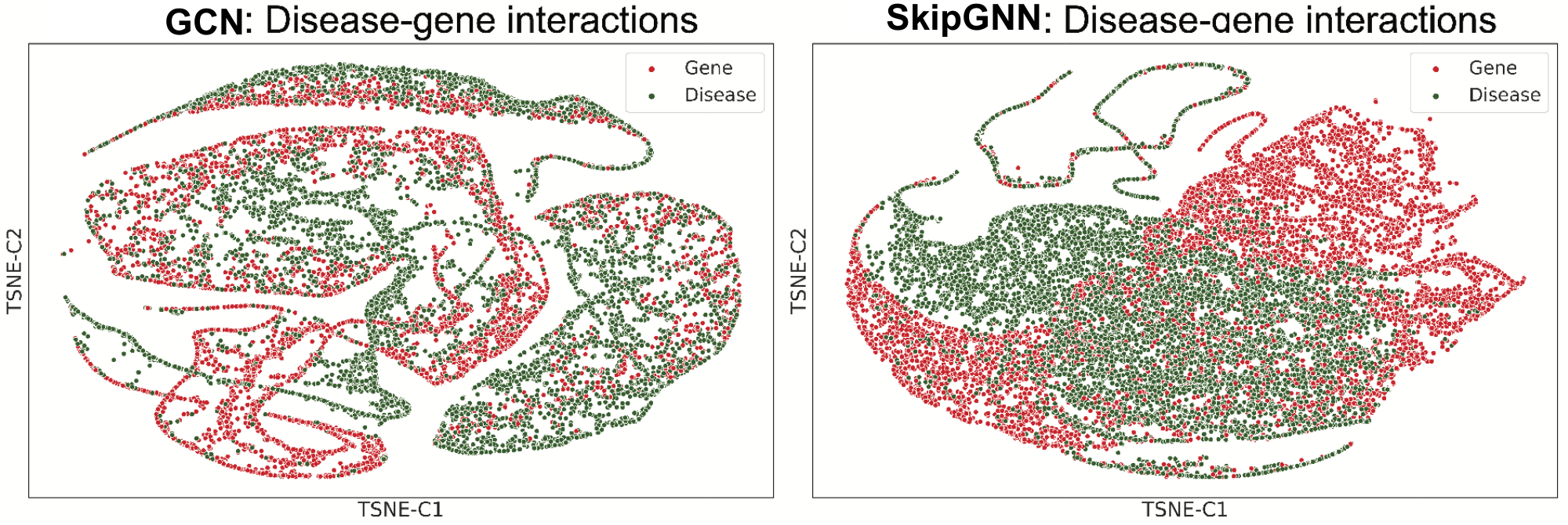}
\caption{{\bf Visualizations of gene-disease interaction network.} GCN does not distinguish disease and gene as it only captures direct similarity whereas \mname is able to distinct disease and gene embeddings, confirming its ability to capture \textit{skip similarity}. We use GCN and \mname on the gene-disease interaction dataset to learn gene/disease embeddings, which are visualized using t-SNE.}
\label{fig:explain-gdi}
\end{figure}

Next, we visualize embeddings learned by GCN and \mname in an effort to investigate whether \mname can better capture the structure of interaction networks than GCN. For that, we use DTI and GDI networks in which drugs/diseases are linked to associated proteins/genes. We use t-SNE~\cite{maaten2008visualizing} and visualize the learned embeddings in Figure~\ref{fig:explain-dti} (DTI network) and Figure~\ref{fig:explain-gdi} (GDI network). Note that both GCN and \mname uses the same input embedding, which means the only difference is whether or not skip similarity is used.

First, we observe that GCN cannot distinguish between different types of biomedical entities (\ie, drugs vs. proteins and disease vs. genes). In contrast, \mname can successfully separate the entities, as evidenced by distinguishable groups of points of the same color in the t-SNE visualizations. This observation confirms that \mname has a unique ability to capture the \textit{skip similarity} whereas GCN cannot. This is because GCN forces embeddings of connected drug-protein/gene-disease pairs to be similar and thus it embeds those pairs close together in the embedding space. However, by doing so, GCN conflates drugs with proteins and genes with diseases. In contrast, \mname generates a biologically meaningful embedding space in which drugs are distinguished from proteins (or, genes from diseases) while drugs are still positioned in the embedding space close to proteins to which they bind (or, in the case of GDI network, diseases are positioned close to relevant disease-associated genes). 

{\color{black} We also calculate the silhouette score of the t-SNE plot, which measures the inter-cluster and intra-cluster distance and is used to calculate the goodness of a clustering technique. A higher value indicates that the sample is better matched to its own cluster and poorly matched to neighboring clusters. Here \mname has a silhouette score of 0.114 for DTI whereas GCN has a score of 0.014 for DTI. For GDI, SkipGNN has a score 0.079 and GCN has a score 0.018. The up to 8 times increase in silhouette scores suggest that SkipGNN can better distinguish the entities than GCN.}

Further, we find that GCN and its graph convolutional variants cannot capture \textit{skip similarity} because they aggregate neural messages only from direct (\ie, immediate) neighbors in the interaction network. \mname solves this problem by passing and aggregating neural message from direct as well as in-direct neighbors, thereby explicitly capturing \textit{skip similarity}. 

\subsection{Ablation studies} \label{sec:exp4}

\begin{table}
\centering

\caption{\label{tab:q3}{\bf Results of ablation experiments.} \mname~'s model components setup achieve the best result. Ablation study result of five independent runs on DDI, PPI and DTI tasks.} 

\begin{tabular}{l|lcc}
\toprule 
Task & Method & PR-AUC & ROC-AUC \\ \midrule
    \multirow{5}{*}{DTI}
    & \mname~& $\bf{0.928 \pm 0.006}$ & $0.922 \pm 0.004$ \\
    \cline{2-4}
    & -fusion &$ 0.909 \pm 0.011$ &$ 0.907 \pm 0.013 $ \\
    & -skipGraph &$0.904 \pm 0.011$ &$0.899 \pm 0.010$ \\
    & -Weighted-L1 & $0.927 \pm 0.013$ & $\bf{0.926 \pm 0.011}$  \\
    & -Hadamard & $0.796 \pm 0.116 $ & $0.795 \pm 0.116$ \\
    \midrule
    \multirow{5}{*}{DDI}
    & \mname~& $\bf{0.866 \pm 0.006}$ &$\bf{0.886 \pm 0.003}$ \\ 
    \cline{2-4}
    & -fusion &$ 0.864 \pm 0.007$& $ 0.884 \pm 0.002$ \\
    & -skipGraph & $0.856 \pm 0.005$ &$0.875 \pm 0.004$  \\
    & -Weighted-L1 & $0.863 \pm 0.006$ &$0.885 \pm 0.003$ \\
    & -Hadamard &$0.833 \pm 0.054$ & $0.883 \pm 0.003$\\
    \bottomrule
    \end{tabular}
    \quad
    \begin{tabular}{l|lcc}
    \toprule 
    Task & Method & PR-AUC & ROC-AUC \\ \midrule

    \multirow{5}{*}{PPI}
    & \mname~& $\bf{0.921 \pm 0.003}$ &$\bf{0.917 \pm 0.004}$  \\
    \cline{2-4}
    & -fusion & $0.912 \pm 0.004$ & $0.906 \pm 0.005$\\
    & -skipGraph  &$0.909 \pm 0.002$ &$0.907 \pm 0.006$  \\
    & -Weighted-L1 & $0.917 \pm 0.003$ &$0.908 \pm 0.006$  \\
    & -Hadamard & $0.909 \pm 0.025$ & $0.914 \pm 0.010$ \\
    \midrule
    \multirow{5}{*}{GDI}
    & \mname~& $\bf{0.915 \pm 0.003}$ &$\bf{0.912 \pm 0.004}$ \\
    \cline{2-4}
    & -fusion & $0.896 \pm 0.029$ & $0.892 \pm 0.014$\\
    & -skipGraph &$0.909 \pm 0.002$ &$0.906 \pm 0.006$ \\
    & -Weighted-L1 & $0.913 \pm 0.009$ &$0.898 \pm 0.010$  \\
    & -Hadamard & $0.883 \pm 0.041$ & $0.891 \pm 0.025$ \\
    \bottomrule
\end{tabular}
\end{table}

To show that each component of \mname has an important role in the final performance of \mname, we conduct a series of ablation studies. \mname has four key components, and we study how the metho performance changes when we remove each of the components:
\begin{itemize}
    \item \textbf{-fusion} replaces \mname's fusion scheme with a simple concatenation of node embeddings generated by GCN. 
    \item \textbf{-skipGraph} removes skip graph and degenerates to GCN.
    \item \textbf{-Weighted-L1} uses weighted-L1 gate in Eq.~(\ref{eq:fuse-gate}) as $\mathrm{AGG}(A, B) = \vert A-B \vert$, where $\vert \cdot \vert$ is the absolute value operator.
    \item \textbf{-Hadamard} replaces the summation gate with Hadamard operator `$*$` in Eq.~(\ref{eq:fuse-gate}) such that $\mathrm{AGG}(A,B) = A * B$.
\end{itemize}
Table~\ref{tab:q3} show results of deactivating each of these components, one at a time. We find that -fusion outperforms -skipGraph (\ie, GCN) by a large margin. This finding identifies skip graph as a key driver of performance improvement. Further, we find that our iterative fusion scheme is important, indicating that successful methods need to integrate both direct and \textit{skip similarity} in interaction networks. Next, we see that weighted $L_1$ gate has comparable or worse performance than the summation gate and Hadamard operator performs the worst, suggesting that \mname's summation gate is the best-performing aggregation function. Altogether, we conclude that all \mname's components are necessary for its strong performance.

\subsection{Investigation of \mname's novel predictions}\label{sec:exp5}
\begin{table}
\centering
\caption{\label{tab:novel_hits} {\bf Novel predictions of drug-drug interactions.} Shown are top-10 predicted drug-drug interactions together with the relevant literature providing evidence for predictions.}
\begin{tabular}{llll}
\toprule     
Rank & Drug 1 & Drug 2 & Evidence for DDI \\ \midrule
     1 & Warfarin & Clozapine & Mukku et al., 2018 \cite{mukku2018clozapine} \\
     2 & Warfarin & Ivacaftor & Robertson et al., 2015 \cite{robertson2015clinical} \\
     3 & Phenelzine & Deferasirox & \\
     4 & Warfarin & Paraldehyde & DuPont, Product Information \cite{warfarin} \\
     5 & Warfarin & Cyclosporine & Snyder, 1988 \cite{cyclo} \\
     6 & Phenytoin & Sipuleucel-T & \\
     7 & Warfarin & Netupitant &  \\
     8 & Phenelzine & Suvorexant& Merck, Product Information \cite{belsomra} \\
     9 & Leuprolide & Picosulfuric acid &  \\
     10 & Deferasirox & Bexarotene & Ligand, Product Information \cite{bexarotene}\\
    \bottomrule
\end{tabular}
\end{table}

The main goal of link prediction on graphs is to find novel hits that do not exist in the dataset. We conduct a literature search and find \mname is able to discover novel hits. We select pairs that are not interacted in the original dataset but are flagged as interaction from our model. We then pick the top 10 confident interactions and feed them into literature database and see if there are evidence supporting our findings. We find promising result for the DDI task (Table~\ref{tab:novel_hits}). Out of the 10 top-ranked interaction pairs, we are able to find 6 pairs that have literature evidence support.

For example, for the interaction between Warfarin and Calozapine, \cite{mukku2018clozapine} reports that ``\textit{Clozapine} increase the concentrations of commonly used drugs in elderly like digoxin, heparin, phenytoin and \textit{Warfarin} by displacing them from plasma protein. This can lead to increase in respective adverse effects with these medications.'' Also, the manufacturer~\cite{clozapine} also reports that ``\textit{Clozapine} may displace \textit{Warfarin} from plasma protein-binding sites. Increased levels of unbound \textit{Warfarin} could result and could increase the risk of hemorrhage.'' Take another example between Warfarin and Ivacaftor, \cite{robertson2015clinical} conducts a DDI study and reports that ``caution and appropriate monitoring are recommended when concomitant substrates of CYP2C9, CYP3A and/or P‐gp are used during treatment with \textit{Ivacaftor}, particularly drugs with a narrow therapeutic index, such as \textit{Warfarin}.'' Finally, we provide the top 10 outputs for DTI, PPI, and GDI tasks in Appendix~\ref{appendix:novel_hits}.

\section{Discussion}

We introduced \mname, a novel graph neural network for predicting molecular interactions. The architecture of \mname is motivated by a principle of connectivity, which we call \textit{skip similarity}. Remarkably, we found that skip similarity allows \mname to much better capture structural and evolutionary forces that govern molecular interaction networks that what is possible with current graph neural networks. \mname achieves superior and robust performance on a variety of key prediction tasks in interaction networks and performs well even when networks are highly incomplete. 

There are several future directions. We focused here on networks in which all edges are of the same type. As \mname is a general graph neural network, it would be interesting to adapt \mname to heterogeneous networks, such as drug-gene-disease networks. Another fruitful direction would be to implement skip similarity in other types of biological networks.

\bibliography{sample}

\section{Author contributions statement}

K.H., C.X., J.S. conceived the projects, K.H., C.X., M.Z., J.S. conceived the experiments, K.H. conducted the experiments. All authors analyzed the results and reviewed the manuscript. 

\section{Additional information}

\textbf{Competing interests} 
The authors declare no competing interests.

\appendix

\section{Experiments on the importance of each layer of GNN for biomedical link prediction}

To further support our claim on the importance of integrating skip similarity for GNN-based methods on biomedical interaction network link prediction, we vary the architecture of vanilla GNN and perform predictive comparison on DDI, PPI, and DTI tasks. Here are the variations:
\begin{itemize}
    \item \textbf{TwoLayers-OriGraph} is the two layers GCN on original graph. It uses an indirect two-hops neighborhood aggregation because the two-hops nodes information is conveyed to the center node through the one-hop nodes.
    \item \textbf{OneLayer-OriGraph} is a one layer vanilla GCN. It only utilizes the immediate one-hop neighbor information. Hence, it is a direct measure of \textit{direct similarity}.
    \item \textbf{TwoLayers-SkipGraph} is the vanilla two layers GCN that operates on the skip graph. It uses direct connection of center node with its two-hops neighborhood as against the indirect connection in vanilla GCN. As it is two layer, it also considers indirect four-hops neighbor nodes.
    \item \textbf{OneLayer-SkipGraph} is the one layer version of GCN-A2. As it only uses two-hop neighbor information, it directly measures the \textit{skip similarity}.
    \item {\color{black} \textbf{OneLayer-3Hops} is the one layer version of GCN-A3. We test to show the significance of higher order neighbors. }
\end{itemize}

Table~\ref{tab:q2} compares the results. From the large improvement of TwoLayers-OriGraph over OneLayer-OriGraph, this is the initial evidence that two-hops neighborhood, which contains \textit{skip similarity} node relation assumption, is essential. Then, comparing OneLayer-OriGraph and OneLayer-SkipGraph, the large margin improvement of OneLayer-SkipGraph implies two-hops neighbor alone has more predictive information than one-hop neighbor alone, supporting our motivation analysis of the importance of \textit{skip similarity} for biomedical interaction network. Note also that the improvement from OneLayer-OriGraph to TwoLayers-OriGraph is much larger than the improvement from OneLayer-SkipGraph to TwoLayers-SkipGraph, meaning second-hop is essential and higher-order neighborhood is of limited importance for interaction link prediction. Lastly, TwoLayers-OriGraph performs better than TwoLayers-SkipGraph, meaning that biomedical interaction link prediction is a balance between immediate neighbor and two-hops neighbor, confirming with our intuition that an ideal network should pursue a balance between them and adding support for the iterative fusion scheme. {\color{black} Note that OneLayer-SkipGraph uses only the second hop neighborhood, without the first-hop neighborhood information. This suggests first-hop importance, and a necessity to integrate both first and second hop neighbors, such as SkipGNN's iterative fusion scheme. We also find 3-hops neighbor is less important than 2-hops neighbor when comparing OneLayer-SkipGraph and OneLayer-3Hops, further confirming the importance of 2-hops in biomedical interaction network.}

\begin{table}
\centering
\color{black}

\caption{\label{tab:q2}{\bf \textit{Skip Similarity} is important for biomedical interaction prediction when using GCN.} Results of five independent runs on DDI, PPI and DTI tasks with varying architectures of GCN.} 
\vspace{3mm}
\begin{tabular}{l|lcc}
\toprule 
Task & Method & PR-AUC & ROC-AUC \\ \hline
    \multirow{5}{*}{DTI}
     & TwoLayers-OriGraph  &$0.904 \pm 0.011$ &$0.899 \pm 0.010$  \\
    & OneLayer-OriGraph  & $0.807 \pm 0.024$ &$0.781 \pm 0.026$\\
    & TwoLayers-SkipGraph  & $0.849 \pm 0.041$ &$0.826 \pm 0.052$ \\
    & OneLayer-SkipGraph & $0.780 \pm 0.047$ & $0.756 \pm 0.046$ \\ & OneLayer-3Hops & $0.806 \pm 0.005$ & $0.789 \pm 0.017$ \\
    \hline\hline
    \multirow{5}{*}{DDI}
    & TwoLayers-OriGraph & $0.856 \pm 0.005$ &$0.875 \pm 0.004$  \\
    & OneLayer-OriGraph & $0.810 \pm 0.029$ &$0.831 \pm 0.029$  \\
    & TwoLayers-SkipGraph & $0.848 \pm 0.003$ & $0.863 \pm 0.002$  \\
    & OneLayer-SkipGraph &$0.844 \pm 0.008$ & $0.862 \pm 0.004$  \\
    & OneLayer-3Hops & $0.830 \pm 0.013$ & $0.851 \pm 0.009$ \\
    \hline\hline
    \multirow{5}{*}{PPI}
    & TwoLayers-OriGraph &$0.909 \pm 0.002$ &$0.907 \pm 0.006$    \\
    & OneLayer-OriGraph &$0.806 \pm 0.013$  &$0.815 \pm 0.015$ \\
    & TwoLayers-SkipGraph & $0.900 \pm 0.003$ &$0.888 \pm 0.004$  \\
    & OneLayer-SkipGraph & $0.873 \pm 0.023$ &$0.863 \pm 0.017$  \\
    & OneLayer-3Hops & $0.858 \pm 0.039$ & $0.853 \pm 0.022$ \\
    \hline\hline
     \multirow{5}{*}{GDI}
    & TwoLayers-OriGraph &$0.909 \pm 0.002$ &$0.906 \pm 0.006$ \\
    & OneLayer-OriGraph &$0.846 \pm 0.043$  &$0.845 \pm 0.039$ \\
    & TwoLayers-SkipGraph & $0.888 \pm 0.008$ &$0.905 \pm 0.007$  \\
    & OneLayer-SkipGraph & $0.876 \pm 0.016$ &$0.883 \pm 0.018$  \\
    & OneLayer-3Hops & $0.868 \pm 0.006$ & $0.881 \pm 0.011$ \\\bottomrule
\end{tabular}
\end{table}

\section{Details about baseline methods}

\begin{itemize}
    \item \textbf{L3}~\cite{kovacs2019network} counts the length-3 paths among all the network nodes pairs. The number of length-3 paths are then normalized by the degree of node pairs.
    \item \textbf{DeepWalk}~\cite{perozzi2014deepwalk} performs uniform distributed random walk and applies skip-gram model to learn a node embedding. We use 20 walk lengths and then concatenate the target nodes embedding with a logistic regression classifier.
    \item \textbf{node2vec}~\cite{grover2016node2vec} builds on DeepWalk and uses biased random walk based on depth/breath first search to consider both local and global network structure. We use 20 walk length as the paper suggests longer walk lengths improve the embedding quality. The paper also reported Hadamard product perform better than average and weighted L1/L2 for link prediction. However, in our experiment, the simple concatenation is better than Hadamard. After the concatenation, we feed into a logistic regression classifier as described in the paper.
    \item \textbf{struc2vec}~\cite{ribeiro2017struc2vec} leverages the local network structure in addition to the node2vec. We use 80 walk length and 20 number of walks, following author's recommendation. We then concatenate the latent embedding and feed into a logistic regression classifier.
    \item \textbf{Spectral Clustering}~\cite{tang2011leveraging} projects nodes on top-16 eigenvectors of the normalized Laplacian matrix and uses the transposed eigenvectors as node embeddings. The embeddings are then multiplied and pass through a sigmoid function to obtain link probabilities.
    \item \textbf{VGAE}~\cite{kipf2016variational} applies variational graph auto-encoder and learns node embeddings that best reconstruct the adjacent matrix. We use a two-layer GCN with hidden size 64 for layer one and 16 for layer two. The learning rate is set to be 5e-4 with Adam optimizer for 300 epochs. The dropout rate is set to be 0.1.    
    \item \textbf{GCN}~\cite{kipf2017semi} uses two-layers GCN layers on original adjacency matrix to obtain node embeddings, others are with same setting as \mname~. We use a two-layer GCN with hidden size 64 for layer one and 16 for layer two. The learning rate is set to be 5e-4 with Adam optimizer for 10 epochs with batch size 256. 
    \item {\color{black} \textbf{GIN}~\cite{GIN} uses multi-layer perceptron (MLP) as the aggregation function. We use a five layer GIN with hidden size 32. The learning rate is set to be 5e-4 with Adam optimizer for 10 epochs with batch size 256.} 
    \item {\color{black} \textbf{JK-Net}~\cite{xu2018representation} uses skip connections across each layer of GNN propapagation. We use the GIN backend for JK-Net. We use three layers GIN with hidden size 64. The learning rate is set to be 5e-4 with Adam optimizer for 10 epochs with batch size 256.} 
    \item {\color{black} \textbf{MixHop}~\cite{abu2019mixhop} uses multiple higher-order adjacency matrix to propagate messages. We use three layers for both the top and lower towers with size 200, 200, 200. The L2 regularization is set to be 0.0005. }
\end{itemize}
We determine all parameters for the baseline methods using the random search on a validation set. 

\section{Potential novel hits for PPI, DTI, and GDI}~\label{appendix:novel_hits}

We conducted a literature search for the DDI novel hits in the main text. Here, we also provide the novel hits discovered through \mname for the PPI, DTI, and GDI tasks in Table~\ref{tab:potential_novel_hits}.

\begin{table}

\centering

\caption{\label{tab:potential_novel_hits}{\bf Top-ranked novel predictions for PPI, DTI and GDI tasks.} Potential novel hits for PPI, DTI, GDI tasks. For long drug names, we use the DrugBank ID instead. }

\vspace{2mm}

\begin{tabular}{l|ccc}
\hline     
Task & Rank & Drug & Target Gene \\ \midrule
    \multirow{10}{*}{DTI}
    & 1 & Dpb-T & L3MBTL1  \\
    & 2 & Progabide & PANX1  \\
    & 3 & Glutamic acid & NARS2  \\
    & 4 & DB04530 %S,S-(2-Hydroxyethyl)Thiocysteine
    & CYP2D6 \\
    & 5 & Glutamic acid
 & AZIN2 \\
    & 6 & DB08152 %{(2S)-1-[N-(tert-butoxycarbonyl)glycyl]pyrrolidin-2-yl}methyl (3-chlorophenyl)acetate
 & BCHE \\
    & 7 & CR002 & CYP2C19 \\
    & 8 & Ecabet & F2 \\
    & 9 & Insulin & CYP2C19 \\
    & 10 & RU84687 & CYP2C19 \\
    \hline \hline
    Task & Rank & Protein 1 & Protein 2  \\ \hline
    \multirow{10}{*}{PPI}
    & 1 & GOLGA6A & CYSRT1  \\
    & 2 & CYSRT1 & IPCEF1  \\
    & 3 & PRKAR1B & REL  \\
    & 4 & CEP70 & UBQLN1  \\
    & 5 & ADAMTSL4 & MTUS2 \\
    & 6 & SIX1 & CYSRT1 \\
    & 7 & RBPMS & MTUS2 \\
    & 8 & KRT31 & CCDC36 \\
    & 9 & TRAF2 & MIPOL1 \\
    & 10 & DES & MEOX2 \\
    \hline \hline
    Task & Rank & Gene & Disease \\ \hline
    \multirow{10}{*}{GDI}
    & 1 & RAPGEF3 & Intellectual disability  \\
    & 2 & ISL1 & Intellectual disability  \\
    & 3 & UHRF1BP1L & Intellectual disability  \\
    & 4 & RCN3 & Intellectual disability \\
    & 5 & UGT1A4 & Schizophrenia \\
    & 6 & DDX11 & Schizophrenia \\
    & 7 & SOD2 & Mental Retardation \\
    & 8 & IL1B & Diabetes mellitus type 2 \\
    & 9 & TNF & Osteosarcoma \\
    & 10 & POMC & Tactile Allodynia \\
    \bottomrule
\end{tabular}
\end{table}

\section{A Network Heuristic Explanation} \label{sec:robustness}

\begin{figure}
    \centering
    \includegraphics[width=0.5\textwidth]{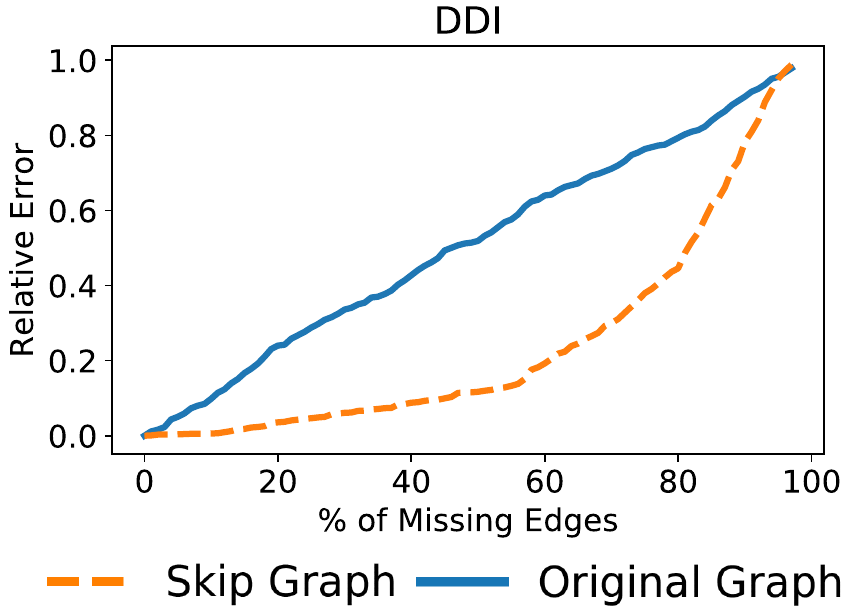}
    \caption{{\bf The ability of skip graph and original graph to capture the network structure in the face of incomplete data.} Skip graph can better preserve the network structure than the original graph, as evidenced by skip graph's smaller relative error (Section~\ref{sec:missing}) than that of the original graph. This is true for all \% of missing edges, indicating that skip graph can keep useful information about interaction structure even when networks are highly incomplete and many interactions are missing.}
\label{fig:spec}
\end{figure}

So far, we found that \mname has robust performance on incomplete interaction networks and next we investigate what makes \mname to perform so robustly. We hypothesize that \mname~is robust because its skip graphs can preserve the graph topology much better than original graphs and this feat becomes prominent when interaction data are scarce. Note that \mname~uses the skip graph whereas other methods only use the original graph. 

To test the hypothesis, we measure the relative error between the original graph $G$ and the incomplete graph $G^{p}$ in which edges are missing at rate $p$. We use a metric that calculates the relative error of the spectral norm for the graph Laplacian matrix: $\mathrm{Err}(\mathbf{A}, p) = (\Vert \mathbf{L}\Vert_2 - \Vert \mathbf{L}^{p} \Vert_2)/\Vert \mathbf{L}\Vert_2,$
where $\mathbf{L} = \mathbf{A} - \mathbf{D}$, $\mathbf{L}^{p} = \mathbf{A}^{p} - \mathbf{D}^{p}$, $\mathbf{A}$ ($\mathbf{A}^p$) is adjacency matrix of $G$ ($G^p$), $\Vert \cdot \Vert_{2} = \sigma_{\max}(\cdot)$, the $\sigma_{\max}$ is the largest singular value~\cite{chung1997spectral}.  

Figure~\ref{fig:spec} shows the relative error $\mathrm{Err}$ of original and skip graphs against 100 fractions $p$ of missing edges on the DDI task. We see that the skip graph's relative error is much lower than that of original graph in almost all settings. This observation provides evidence for our hypothesis, confirming that skip graphs can better capture the graph topology than original graphs. Because of that, \mname~can learn high-quality embeddings even when interaction data are scarce.

\section{{\color{black} Biomedical Interaction Network Visualization}}

\begin{figure}[h]
    \centering
    \includegraphics[width = 0.6 \textwidth]{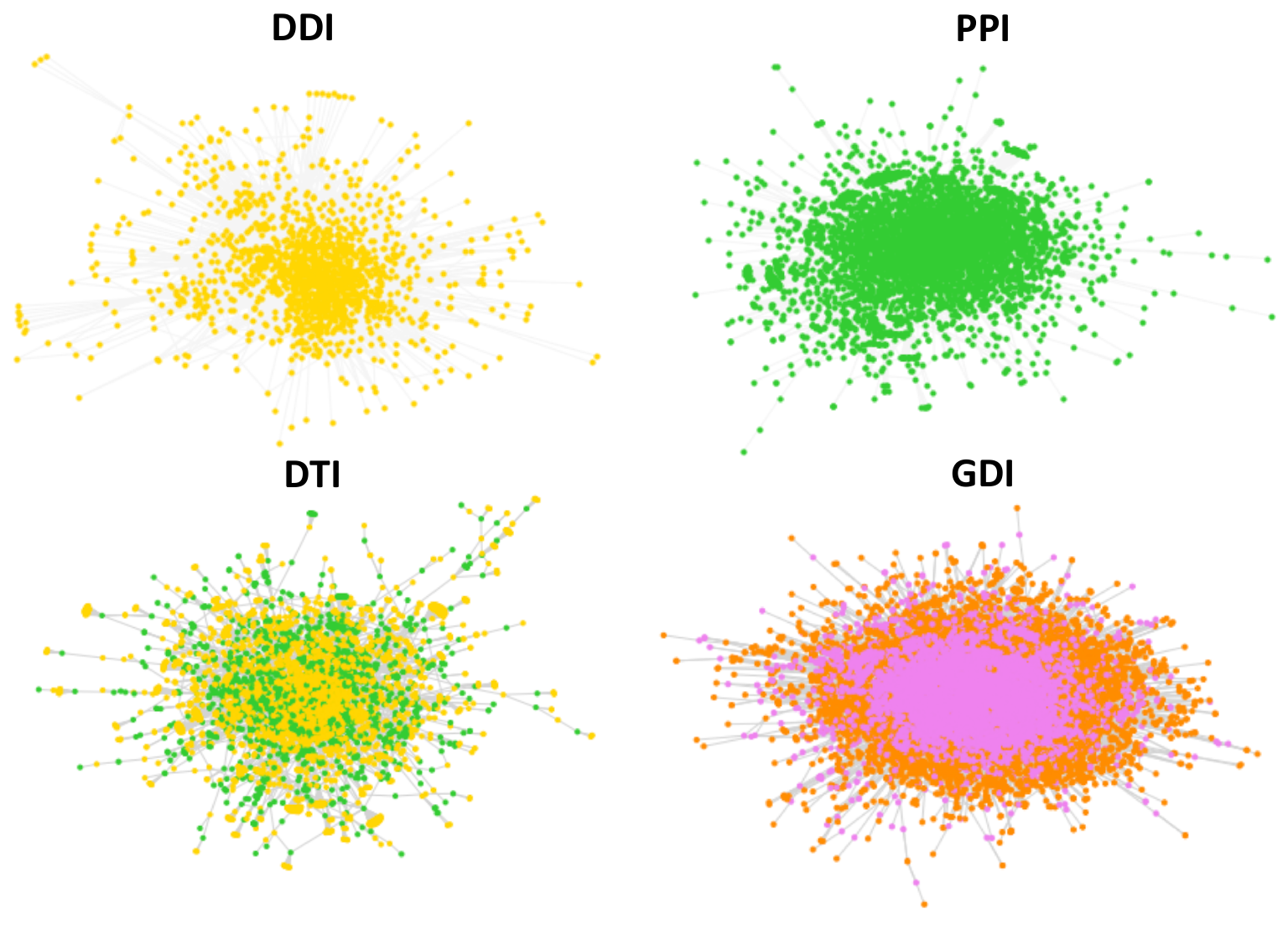}
    \caption{{\color{black}Network Visualization of four biomedical interaction networks.}}
    \label{fig:viz}
\end{figure}
A visualization of biomedical network is provided in Fig.~\ref{fig:viz}.

\end{document}